\documentclass[aps,prb,times,twocolumn,superscriptaddress,a4paper]{revtex4-1}

\usepackage{colortbl}
\usepackage{amsfonts,amsmath,amssymb}
\usepackage[dvips]{graphicx}
\usepackage{hyperref,tikz}
\usepackage{tabularx,hhline}
\usepackage{float} 
\usepackage{dcolumn,stmaryrd}

\hypersetup{colorlinks=true,linkcolor=blue,citecolor=red, urlcolor=black}

\begin{document}

\title{Monopole holes in a partially ordered spin liquid}
\author{L.D.C. Jaubert}
\affiliation{Okinawa Institute of Science and Technology Graduate University,
Onna-son, Okinawa 904-0495, Japan}

\date{\today}
\begin{abstract}
If spin liquids have been famously defined by what they are not, \textit{i.e.} ordered, the past years have seen the frontier between order and spin liquid starting to fade, with a growing number of materials whose low-temperature physics cannot be explained without co-existence of (partial) magnetic order and spin fluctuations. Here we study an example of such co-existence in the presence of magnetic dipolar interactions, related to spin ice, where the order is long range and the fluctuations support a Coulomb gauge field. Topological defects are effectively coupled via energetic and entropic Coulomb interactions, the latter one being stronger than for the spin-ice ground state. Depending on whether these defects break the divergence-free condition of the Coulomb gauge field or the long-range order, they are respectively categorized as monopoles -- as in spin ice -- or \textit{monopole holes}, in analogy with electron holes in semiconductors. The long-range order plays the role of a fully-occupied valence band, while the Coulomb spin liquid can be seen as an empty conducting band. These results are discussed in the context of other lattices and models which support a similar co-existence of Coulomb gauge field and long-range order. We conclude this work by explaining how dipolar interactions lift the spin liquid degeneracy at very low energy scale by maximizing the number of flippable plaquettes, in light of the equivalent quantum dimer model.
\end{abstract}

\maketitle


The possibility to recast the collective behavior of electrons and atoms  as elegant emergent phenomena is probably one of the most fascinating aspect of condensed matter. The emergence of quasi-particles does not only allow for a deeper understanding of the problem at hand but has also often led to surprising connections across physics, as recently illustrated by photon-like magnetic excitations in quantum spin ice~\cite{Hermele04a,Banerjee08a,Benton12a,Kato15a}. Such approach takes an enhanced flavor when the particle has not yet been observed at high energy, or may not exist. This is for example the case for Majorana fermions observed in nanowires coupled to superconductors~\cite{Mourik12a}, and for magnetic monopoles and their underlying Coulomb gauge field~\cite{Castelnovo08a}.

Coulomb gauge theories have emerged from a variety of discrete models, on the kagome lattice~\cite{Moessner03a}, in fully-packed loop models~\cite{Jaubert11a}, itinerant-electron systems at partial-filling~\cite{Fulde02a,McClarty14c,Chen14a} and chemically disordered materials such as CsNiCrF$_{6}$~\cite{Banks12a}. We refer the interested reader to the excellent review by Chris Henley on this topic~\cite{Henley10a}. But its most famous experimental realization has probably been observed in the classical spin liquid ground state of spin ice materials Dy$_{2}$Ti$_{2}$O$_{7}$ and Ho$_{2}$Ti$_{2}$O$_{7}$, where topological excitations take the form of magnetic monopoles effectively interacting via long-range Coulomb interactions.

Interestingly, a Coulomb phase is not incompatible with partial magnetic order. For example, the stability of a Coulomb ferromagnet has been discussed in the context of  the generalized Quantum Spin Ice model~\cite{Savary12a,Hao14a} and observed over a finite temperature window in a classical spin ice model with broken rotational symmetry~\cite{Powell15a}. In two dimensions, such co-existence has been studied in the two-stage ordering process of the dipolar kagome ice model~\cite{Moller09a,Chern11a}, of direct interest for artificial magnetic lattices~\cite{Nisoli13a}. In higher dimensions, advanced numerical simulations of the nearest-neighbour valence-bond wave functions on the cubic and diamond lattices have shown persistence of a small but clearly finite squared staggered moment in the thermodynamic limit~\cite{Albuquerque12a}.

Our goal in this paper is to study what happens when a Coulomb spin liquid co-exists with long-range order on the pyrochlore lattice~\cite{Borzi13a,Brooks14a,Guruciaga14a}. In particular, based on the Helmholtz decomposition of Fig.~\ref{fig:half}, we explain in section~\ref{semiconductor} how quasi-particles which break the long-range order can be understood as \textit{monopole holes} in analogy with electron holes in semiconductors. In that sense, the Coulomb spin liquid plays the role of a conduction band where defects are ``standard'' monopoles, while the long-range ordered part of the spin degrees of freedom serves as a fully occupied valence band.\\

After a brief presentation of the physics of spin ice in section~\ref{SI}, we introduce the concept of a Fragmented Coulomb Spin Liquid (FCSL) in section~\ref{FCSL} which will be the central theme of our paper. In particular, the topology of the FCSL and its mapping onto the hard-core dimer model on the diamond lattice are discussed, before presenting the numerical methods used in this work. The nearest-neighbour version of the FCSL is briefly studied in section~\ref{entropic} in order to compute the strength of the \textit{entropic} Coulomb potential between topological defects. Magnetic dipolar interactions are included in section~\ref{hole} and the FCSL is explained in terms of the dumbbell model~\cite{Castelnovo08a}, focusing on the properties due to the underlying co-existence with long-range order. Notably, topological defects can be divided into two kinds of quasi-particles and are found to interact via an effective \textit{energetic} Coulomb potential. Based on these results, section~\ref{semiconductor} is devoted to the similitudes between the FCSL and semiconductor physics. The relevance of our work to other systems, especially on the cubic lattice, is discussed in section~\ref{cubic}. To conclude, in section~\ref{R}, we confirm that dipolar interactions order the FCSL at very low temperature~\cite{Borzi13a} whose ground state we identify to be the so-called R-states that have been studied in the quantum dimer model on the diamond lattice~\cite{Bergman06a,Sikora09a,Sikora11a}. Additionally, the pinch points in the structure factor of Fig.~\ref{fig:SQ} are found to persist down to the transition temperature, despite the build-up of R-states correlations.


\section{What is the Coulomb spin liquid in spin ice ?}
\label{SI}

\begin{figure}[b]
\centering\includegraphics[width=7cm]{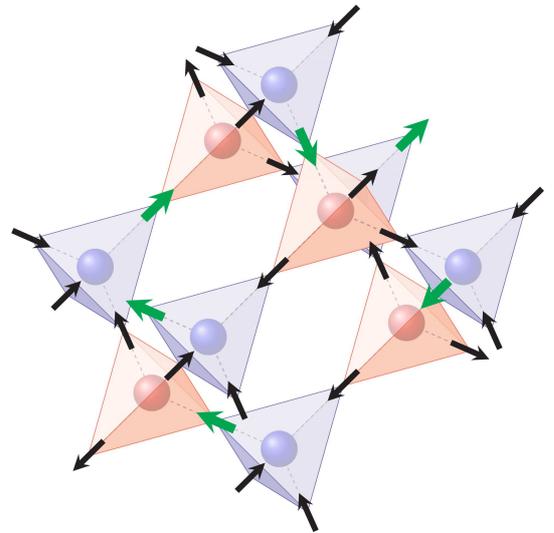}
\caption{The Fragmented Coulomb Spin Liquid (FCSL) studied here is made of Ising spins on the pyrochlore lattice where all up (blue) and down (red) tetrahedra respectively carry effective positive (``3 in - 1 out'') and negative (``3 out - 1 in'') magnetic charges. The magnetic charges crystallize in a zinc-blende structure on the diamond lattice, whose sites are in the centers of the tetrahedra. Time-reversal symmetry reverses the sign of all charges, but conserves the FCSL ensemble. The minority spins are colored in green and shared by both neighbouring tetrahedra. A given spin configuration is entirely determined by the minority spins, modulo time-reversal symmetry, which provides a two-to-one mapping onto a hard-core dimer model on the diamond lattice~\cite{Nagle66c,Bergman06a,Sikora11a}. 
Please note that the fully saturated configuration in a [111] magnetic field is part of the FCSL ensemble.}
\label{fig:CSL}
\end{figure}

The spin ice model consists of Ising spins $\vec S_{i}$ on the pyrochlore lattice made of corner-sharing tetrahedra. The pyrochlore structure is given in Fig.~\ref{fig:CSL}. Each spin belongs to two tetrahedra and can only point towards the center of these two tetrahedra. In spin-ice materials, this constraint is imposed by the surrounding crystal field of the oxygens. Thus for a given tetrahedron, its four spins are not collinear and form four distinct sublattices. For nearest-neighbour ferromagnetic interactions, the ground state of this model is a highly degenerate classical spin liquid~\cite{Harris97a,Ramirez99a} where all tetrahedra have two spins pointing in and two spins pointing out. The resulting coarse-grained magnetization field is locally conserved and divergence-free, conferring the name of Coulomb spin liquid to the ground state of spin ice, in analogy to Maxwell's equations. Experimentally, the resulting spin-spin correlations are dipolar-like and take the form of pinch-points in the structure factor, as measured by neutron scattering in Ho$_{2}$Ti$_{2}$O$_{7}$~\cite{Tabata06a,Fennell07a,Fennell09a,Kadowaki09a}.

But what happens beyond nearest-neighbours ? The question is especially relevant to magnetic dipolar interactions (see Eq.~(\ref{eq:dip})), which can be important for rare-earth ions where nearest-neighbour couplings are rather weak ($\sim 1$K) and single-ion magnetic moments very large (up to $\sim 10 \mu_{B}$):
\begin{eqnarray}
E_{dip}= D r_{p}^{3}\;\sum_{i>j} \frac{\vec S_{i}\cdot \vec S_{j} \, -\, 3 \left(\vec S_{i}\cdot \vec e_{ij}\right)\left(\vec S_{j}\cdot \vec e_{ij}\right)}{r_{ij}^{3}}
\label{eq:dip}
\end{eqnarray}
where $i,j$ are pyrochlore sites separated by a distance $r_{ij}$ along the unit-length vector $\vec e_{ij}$ and $r_{p}$ is the nearest-neighbour distance on the pyrochlore lattice. The size of the magnetic moment $\mu$ is included in the prefactor
\begin{eqnarray}
D =\frac{\mu_{0}\;\mu^{2}}{4\pi\,r_{p}^{3}}
\label{eq:D}
\end{eqnarray}
where $\mu_{0}$ is the vacuum magnetic permeability. Remarkably, the Coulomb-spin-liquid degeneracy is only weakly lifted by such long-range interactions~\cite{Gingras01a,Isakov05a}, because magnetic dipolar interactions between spins can be recast as effective magnetic Coulomb interactions between topological excitations, namely the magnetic monopoles~\cite{Castelnovo08a,Castelnovo10d,Castelnovo12a}. The spin-ice ground state thus appears as a vacuum of magnetic charges. Following the procedure developed in Ref.~[\onlinecite{Castelnovo08a}], the Coulomb potential between two magnetic charges at distance $r$ is
\begin{eqnarray}
V=-\frac{\mu_{0}\,Q^{2}}{4\pi\, r}\quad\textrm{where}\quad Q=\pm \frac{2\mu}{r_{d}} 
\label{eq:dumbbell}
\end{eqnarray}
where $r_{d}$ is the nearest-neighbour distance on the diamond lattice and shortest distance between charges. Using Eq.~(\ref{eq:D}), the potential energy can be expressed dimensionless as follow
\begin{eqnarray}
\frac{V}{D}= -\frac{8}{3}\sqrt{\frac{2}{3}}\;\frac{1}{r/r_{d}}
\label{eq:Vpot}
\end{eqnarray}

At very low energy scales, the quasi-degeneracy of the Coulomb spin liquid is ultimately lifted by dipolar interactions~\cite{Siddharthan99a,Hertog00a,Melko01a} and leads to exotic physics in a magnetic field, such as the magnetisation plateau in a [001] field analogue to a quantum solid phase in $(2+1)$ dimensions~\cite{Lin13b}.\\

\section{What is a \textit{fragmented} Coulomb spin liquid ?}
\label{FCSL}

\subsection{Presentation}

\begin{figure}[t]
\centering\includegraphics[width=8cm]{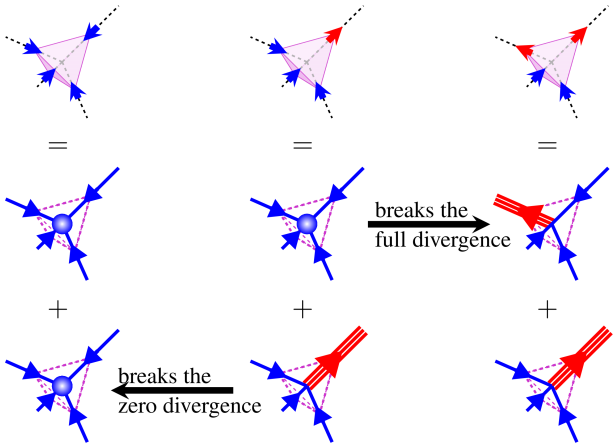}
\caption{The Helmholtz decomposition of the spin degrees of freedom is done as follow~\cite{Brooks14a}. Each spin contributes to two magnetization fluxes. \textit{Left}: for the double charges ``4 in'' (or ``4 out''), all fluxes are pointing inwards (or outwards), giving rise to maximum or full divergence. \textit{Right}: for the ``2 in - 2 out'' configurations, there are as many fluxes pointing inwards as outwards, giving rise to a divergence-free field, \textit{i.e.} with local flux conservation. \textit{Center}: for the ``3 in - 1 out'' configurations, the two contributions are noticeably different, being the sum of a full - and a zero - divergence. In the FCSL, the former contribution represents the long-range charge order, while the latter contribution is the Coulomb spin liquid. Creating a ``4 in / 4 out'' or a ``2 in - 2 out'' defect out of the FCSL locally breaks respectively the zero- or the full-divergence.}
\label{fig:half}
\end{figure}

In spin ice, the nearest-neighbour coupling plays the role of an effective chemical potential for monopoles~\cite{Castelnovo08a}. Using chemical pressure~\cite{Zhou12a,Wiebe15a}, it is possible to tune the chemical potential and favor monopoles. Theoretically, if we forbid the presence of ``4 in / 4 out'' tetrahedra, a large negative chemical potential shall order the singly-charged monopoles into a zinc-blende structure~\cite{Borzi13a,Brooks14a}, as depicted on Fig.~\ref{fig:CSL}. In this long-range charge order, all positive charges have four negatively charged nearest neighbours, and vice-versa. As illustrated in Fig.~\ref{fig:half}, the Helmholtz decomposition of the spin degrees of freedom~\cite{Brooks14a} shows that a fraction of the degrees of freedom forms a divergence-full configuration (the zinc-blende charge order) while the remaining part is divergence-free and supports an extensive degeneracy. In this phase, the local divergence condition is thus half-full, or arguably, half-empty. To emphasize this feature, we shall refer to this partially ordered phase as a Fragmented Coulomb Spin Liquid (FCSL). In particular, we use the term ``fragmented'' as a generalization since other phases may present the same physics with various degree of magnetic order (see discussion in section~\ref{cubic}).\\

\begin{figure}[h]
\centering\includegraphics[width=\columnwidth]{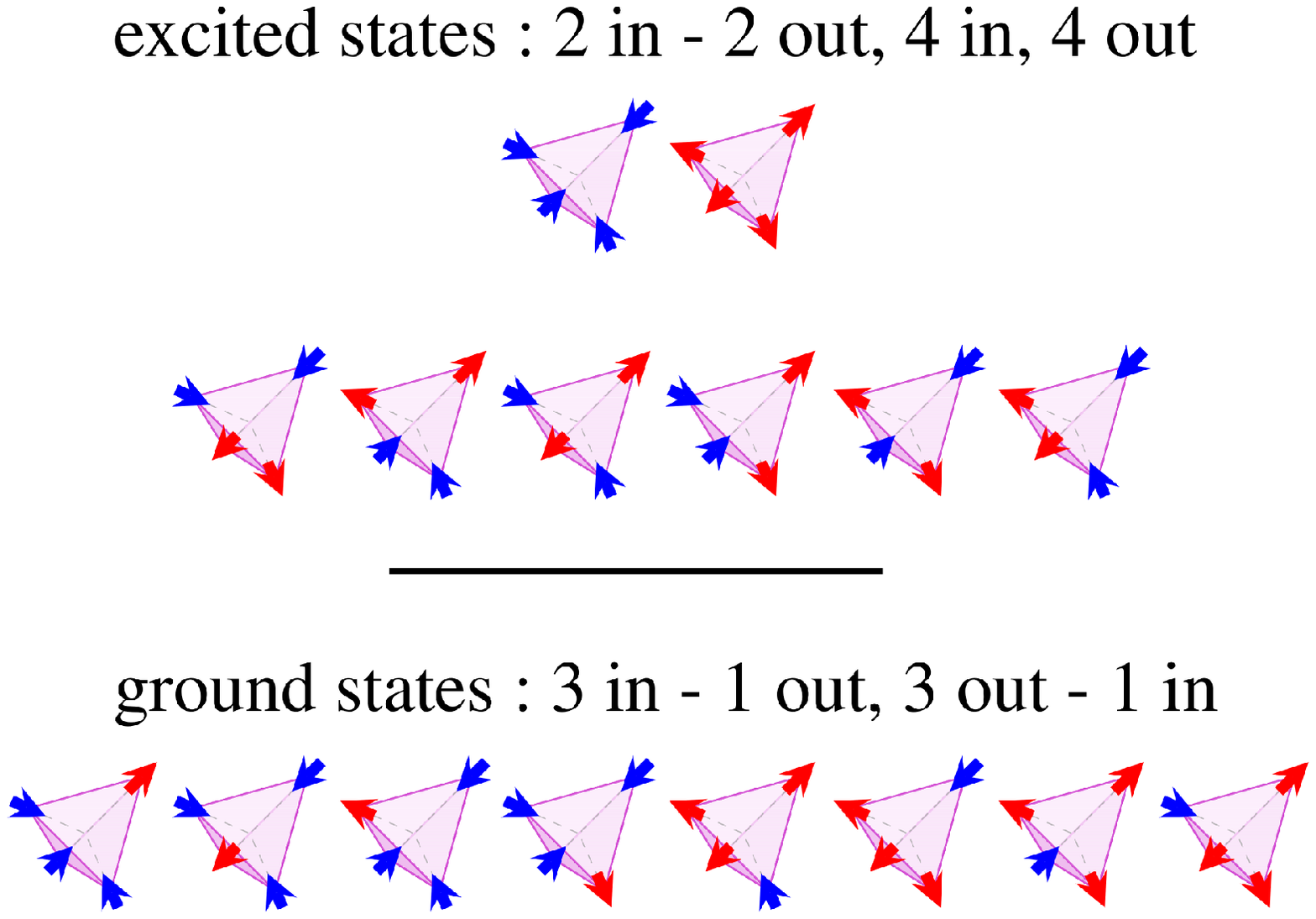}
\caption{
The fragmented Coulomb spin liquid considered in this paper is entirely composed of single charges (``3 in - 1 out'' and ``3 out - 1 in''), as opposed to the spin-ice ground state made of ``2 in - 2 out'' tetrahedra. The ``2 in - 2 out'' and ``4 in / 4 out'' configurations are considered as excitations out of the FCSL. All blue (resp. red) spins are pointing inwards (resp. outwards) of these up tetrahedra.
}
\label{fig:ham}
\end{figure}

The competition between the FCSL and the spin-ice ground state has been studied for fixed density of monopoles~\cite{Borzi13a} and fixed chemical potential~\cite{Brooks14a} in \textit{absence} of ``4 in / 4 out'' tetrahedra. However, in order to study the general properties of the Coulomb spin liquid co-existing with magnetic order, we do not want to eliminate any configuration \textit{a priori}. In presence of  ``4 in / 4 out'' tetrahedra, the FCSL is a priori unstable with only nearest-neighbour Ising spin-spin coupling and dipolar interactions~\cite{Brooks14a,Guruciaga14a}. But it can be stabilized for example via four-body interactions
\begin{eqnarray}
\mathcal{H}=J_{\square}\sum_{\nu}\left(\sum_{\langle ij\rangle \in \nu} \vec S_{i}\cdot\vec S_{j}\right)^{2},
\label{eq:ham}
\end{eqnarray}
where $\nu$ runs over all tetrahedra in the system and $\langle ij\rangle$ represents the six pairs of spins in tetrahedron $\nu$. As illustrated in Fig.~\ref{fig:ham} for Hamiltonian~(\ref{eq:ham}), the eight single charges are in the ground state (``3 in - 1 out'' and ``3 out - 1 in''), while the six ``2 in - 2 out'' and two ``4 in / 4 out'' configurations are excited states. Dipolar interactions would then naturally order the singly-charged monopoles into a zinc-blende structure as in Fig.~\ref{fig:CSL}. Further nearest-neighbour interactions can then easily tune the energy cost of the ``2 in - 2 out'' and ``4 in / 4 out'' configurations. It should also be noted that the FCSL can be partially stabilized for a precise set of parameters of the dipolar spin ice model~\cite{Guruciaga14a}. More generally, singly-charged monopoles may be favored by quantum fluctuations in the context of Yb$_{2}$Ti$_{2}$O$_{7}$~\cite{Applegate12a} and magneto-electric coupling~\cite{Khomskii12a,Jaubert15a}.

However, our goal in this paper is not to focus on a given Hamiltonian or to propose a mechanism to stabilize the FCSL phase, which will be investigated elsewhere~\cite{Udagawa}. Our goal is to describe and understand how the co-existence of order and spin liquid, mediated by dipolar interactions, affect the emergence of topological defects and eventually the ordering of the spin liquid phase itself. From now on, we shall assume that the FCSL is stable and that ``2 in - 2 out'' and ``4 in / 4 out'' spin configurations are excitations out of the FCSL.\\

\subsection{Pseudo-magnetization}

A convenient way to distinguish between the different phases that will be encountered in this work is via the pseudo-magnetization
\begin{eqnarray}
\rho=\frac{1}{N} \left|\sum_{i} \vec S_{i}\cdot\vec e_{i} \right|
=\frac{1}{N} \left|\sum_{i} \sigma_{i} \right|
\label{eq:sigma}
\end{eqnarray}
where $\vec e_{i}$ is the local easy-axis vector pointing out of down tetrahedra. Hence the pseudo-spin $\sigma_{i}=\vec S_{i}\cdot\vec e_{i}=-1$ if the spin points inwards a down tetrahedron, and $+1$ if it points outwards. The ``2 in - 2 out'' spin-ice ground state, FCSL and antiferromagnetic ``all in / all out'' long-range order respectively take the values of $\rho=\{0,1/2,1\}$, both globally for the entire system and locally on each tetrahedron. This quantity will be of particular interest when discussing the semiconductor analogy in section~\ref{semiconductor}.\\


\subsection{The dimer model on the diamond lattice}

The minority spins of the FCSL can be recast as dimers on the diamond lattice, as shown in Fig.~\ref{fig:CSL} where the dimers are the green spins. This mapping allows for a direct estimate of the degeneracy of the FCSL, with approximately $1.3^{N/2}$ configurations~\cite{Nagle66c}, where $N$ is the total number of pyrochlore sites. This value is to be compared to the spin-ice ground state degeneracy, with approximately $1.5^{N/2}$ configurations~\cite{Pauling35a,Nagle66a}. The nearest-neighbour version of our model is thus the classical version of the quantum dimer model on the diamond lattice~\cite{Bergman06a,Sikora09a,Sikora11a}, which has been discussed in the context of the magnetization plateau observed in HgCr$_{2}$O$_{4}$ and CdCr$_{2}$O$_{4}$~\cite{Penc04a,Ueda05a,Bergman06b}. This parallel will be useful in section~\ref{R}.\\


\subsection{Topology}

Starting from a given FCSL configuration and flipping a closed chain made of spins which are alternatively pointing in and out of successive neighboring tetrahedra, automatically creates another FCSL configuration. This is because such a spin update respects both the full- and zero-divergence constraints of the FCSL phase. This closed chain of spins is often better known as a ``worm'' because it is used as an update of the worm algorithm in Monte Carlo simulations. A worm can be local and as small as 6 sites, as illustrated by the thick lines on Fig.~\ref{fig:R}, or on the contrary extensively long and winding around the system with periodic boundary conditions. The ``2 in - 2 out'' and ``4 in / 4 out'' excitations mentioned previously are actually topological defects whose creations, diffusion and annihilations correspond to worm updates.\\

The topology of the FCSL phase is thus strongly reminiscent of the topology of the ``2 in - 2 out'' spin-ice ground state~\cite{Henley10a}, but with one main difference. In a topological phase, two configurations belonging to the same topological sector are connected by non-winding-worm updates. When winding worms are necessary and sufficient to transform one configuration into the other, the two configurations belong to different topological sectors, but to the same \textit{Kempe} sector~\cite{Mohar10a,Cepas11a}. A trivial example is the fully saturated configuration along a given [111] direction, which is a topological sector by itself. When two configurations cannot be related by any worm update, then they are said to belong to different Kempe sectors. In the ``2 in - 2 out'' ensemble, any pair of configurations can be transformed into each other by successive worm updates; there is thus only one Kempe sector. This is not the case for the FCSL. By construction, a worm cannot transform a positive charge into a negative one, or vice-versa. Thus two FCSL configurations connected by time-reversal symmetry cannot be connected by worms. It is easy to show that the FCSL is actually divided into two different Kempe sectors. As a consequence of the time-reversal symmetry, a given Kempe sector is selected by a saturating magnetic field in the [111] direction, while reversing the magnetic field selects the other Kempe sector. It means there is a bijective mapping between the dimer model on the diamond lattice and each of the FCSL Kempe sectors. In absence of any time-reversal symmetry breaking term, the properties of the two Kempe sectors are the same.


\subsection{Numerical methods}

The dipolar energy of Eq.~(\ref{eq:dip}) has been implemented by Ewald summation into classical Monte Carlo simulations, without a demagnetization factor~\cite{Deleeuw80a,Melko04a}. A variation of the worm algorithm~\cite{Melko04a,Brooks14a} has been used to separate topological defects in sections~\ref{entropic} and~\ref{hole}, and to equilibrate the system in section~\ref{R}. In the latter case, acceptance of the worm updates was done via a Metropolis argument. To further help thermalization in section~\ref{R}, parallel tempering was also included~\cite{swendsen86a,geyer91a}.

\section{Entropic Coulomb interactions in the nearest-neighbour model}
\label{entropic}

Even though one of the foci of the present paper is on the influence of long-range dipolar interactions, we shall briefly consider nearest-neighbour physics in this section.

 Due to the underlying divergence-free condition, Coulomb phases are known to support \textit{entropic} dipolar correlations between spins~\cite{Isakov04b,Henley05a,Henley10a}, which can be recast as an entropic interaction between topological defects. In the case of the ``2 in - 2 out'' spin-ice phase, this entropic interaction has been computed analytically and checked numerically by Castelnovo \textit{et al.}~\cite{Castelnovo11a}, and found to be
\begin{eqnarray}
E^{\rm ent}=\alpha_{\rm SI}\frac{k_{B} T}{R}\textrm{,   where } \alpha_{\rm SI}=0.36755
\label{eq:entropicV}
\end{eqnarray}
where $T$ is the temperature and $R=r/r_{d}$ is the dimensionless distance between topological defects in units of the diamond nearest-neighbour distance $r_{d}$. As a consequence, the probability for two topological defects in a Coulomb phase to be separated by a vector $\vec R$ is  independent of the temperature
\begin{eqnarray}
P(\vec R)\propto \exp\left(E^{\rm ent}/k_{B} T\right) = \exp\left(\alpha/|\vec R|\right).
\label{eq:entropicP}
\end{eqnarray}

To measure such distribution is not only the signature of an emergent Coulomb gauge theory but also a way to characterize the Coulomb phase by the value of its parameter $\alpha$. Since the exponential law of Eq.~(\ref{eq:entropicP}) is independent of the temperature, it remains valid in the limit $T\rightarrow 0^{+}$ where we can assume that only \textit{one} pair of topological defects persists. This is why we have performed Monte Carlo simulations to measure the probability distribution for a unique pair of topological defects to be separated by vector $\vec R$, or equivalently, the probability distribution to be separated by distance $R=|\vec R|$, renormalized by the number of diamond sites sitting on a sphere of radius $R$. Technically, starting from a random FCSL configuration, we create a pair of topological defects which we separate using a worm algorithm. At each step, one of the defect is randomly moved to a neighbouring diamond site, keeping the number of defects constant, until the pair of defects is annihilated. The distribution is updated at each step. The process is repeated over 10$^{4}$ closed worms and averaged over 100 independent initial configurations. In order to check the validity of the exponential law over large distances, we considered a system made of 16 millions (1.6 10$^{7}$) sites. To validate our simulations against existing literature, we have also simulated the ``2 in - 2 out'' spin-ice phase.

\begin{figure}[h]
\centering\includegraphics[width=8cm]{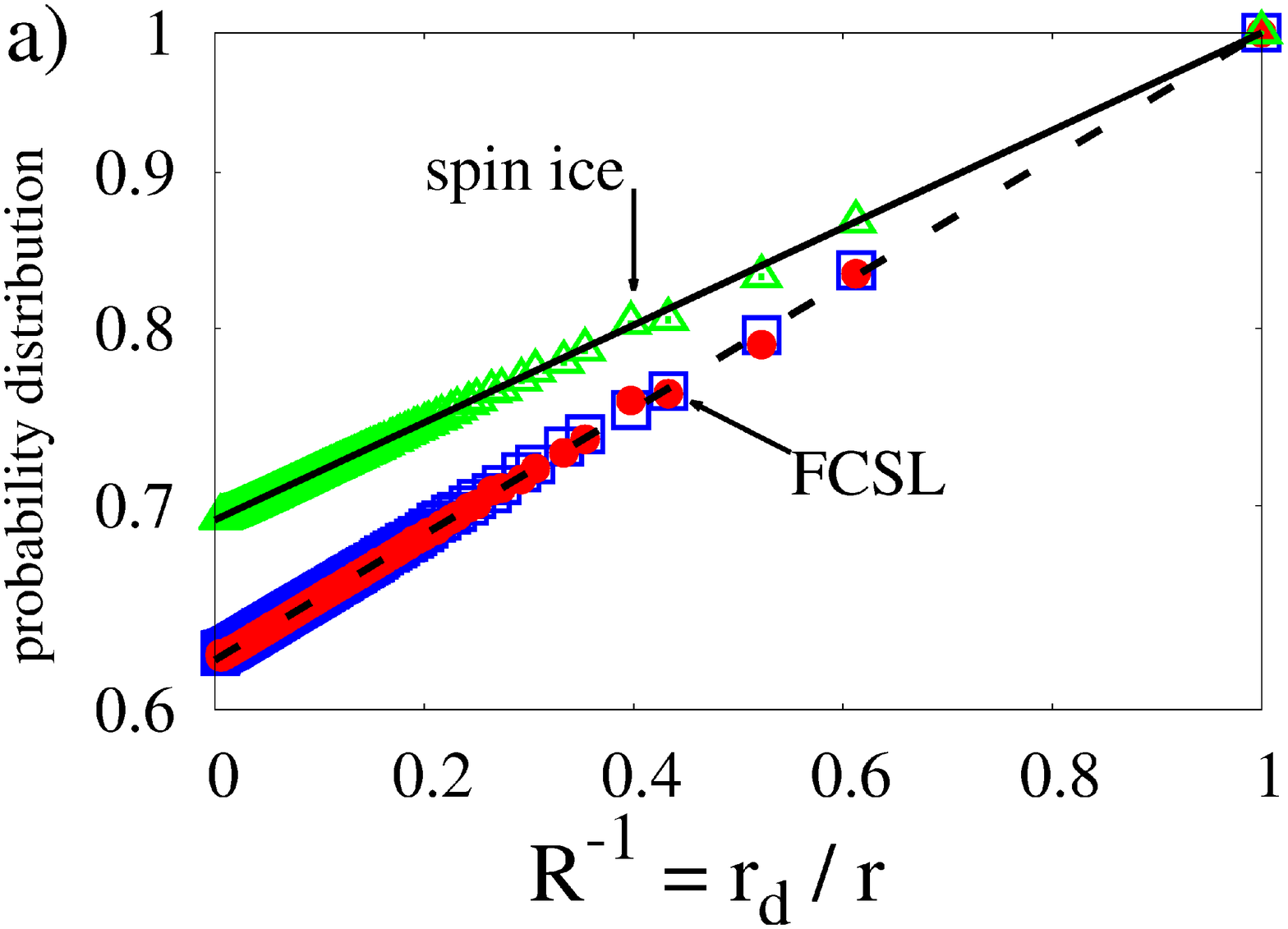}
\centering\includegraphics[width=8cm]{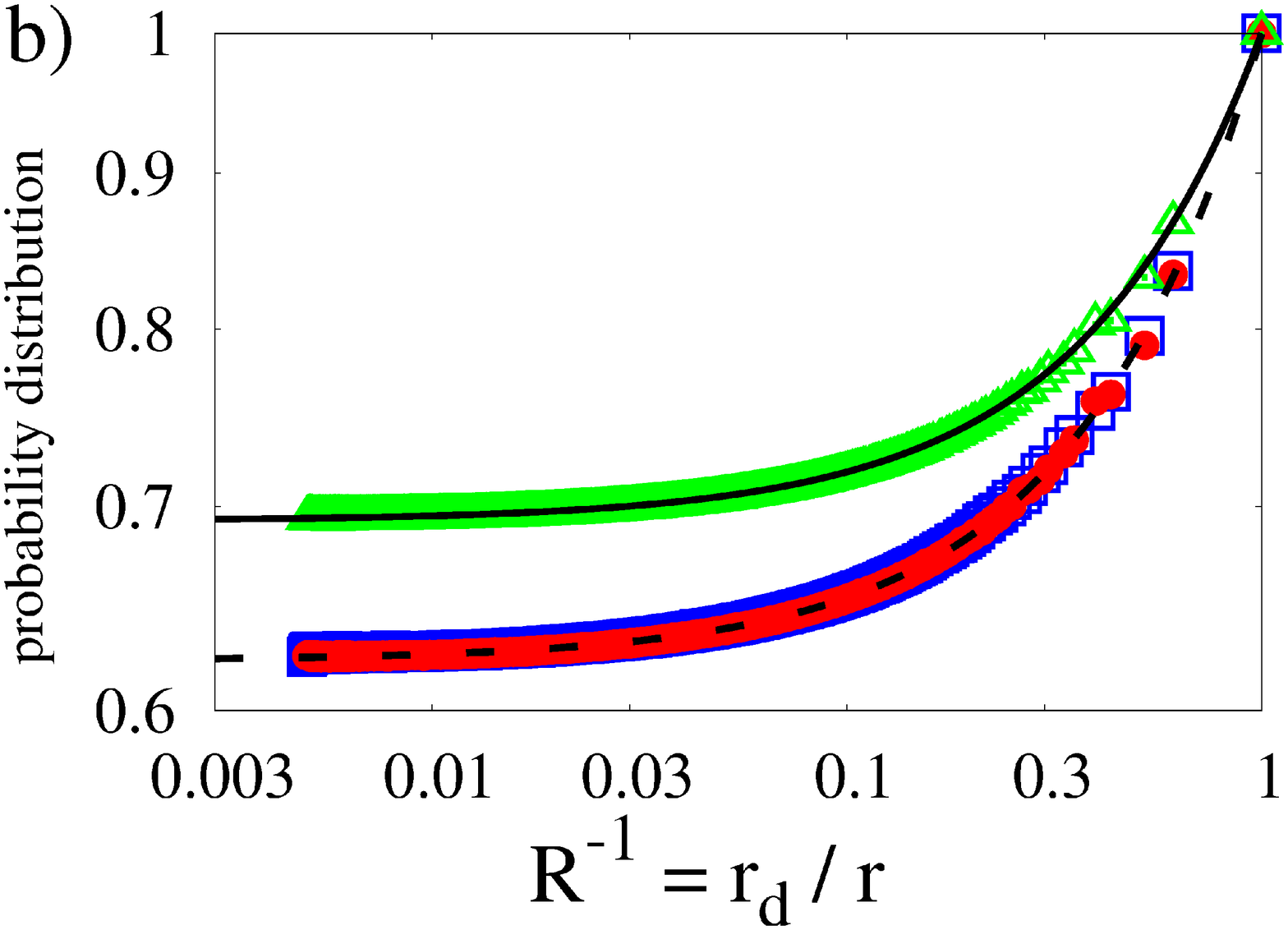}
\caption{
Monte Carlo simulations confirm that, in \textit{absence} of dipolar interactions, the entropic probability distribution for a unique pair of topological defects to be separated by a vector $\vec R$ follows an exponential scaling law as in Eq.~(\ref{eq:entropicP}). This scaling law for the ``2 in - 2 out'' spin-ice phase (black solid line) is known exactly~\cite{Castelnovo11a} and serves as a validity check of our simulations (\textcolor{green}{$\triangle$}). As for the FCSL (\textcolor{blue}{$\square$} and \textcolor{red}{$\bullet$}), the exponential scaling law is found to be characterized by a parameter $\alpha_{\rm FCSL}=0.473\pm0.005$ (dashed line). For the sake of completeness, the two cases where the initial pair of defects out of the FCSL were ``2 in - 2 out'' (\textcolor{red}{$\bullet$}) and ``4 in / 4 out'' (\textcolor{blue}{$\square$}) were considered separately, but no pertinent difference was noticed. The $y-$axes are on a logarithmic scale, while the $x-$axes are on a a) linear and b) logarithmic scale to respectively emphasize the exponential scaling law and the long distance behavior. All data have been arbitrarily normalized to 1 at $R=1$, $r_{d}$ being the shortest distance between charges. The error bars are smaller than the data symbols.
}
\label{fig:entropicP}
\end{figure}

As shown in Fig.~\ref{fig:entropicP}, our numerical results for spin ice (green triangles) perfectly match the theoretical value of $\alpha_{\rm SI}=0.36755$ (solid black line). As for the FCSL, we find a higher entropic interaction of $\alpha_{\rm FCSL}=0.473\pm0.005$. For the sake of completeness, we have considered separately the cases where the initial pair of defects out of the FCSL were ``2 in - 2 out'' and ``4 in / 4 out'', but the parameter $\alpha_{\rm FCSL}$ was found to be the same in both cases. The stiffness of a Coulomb phase is inversely related to the variance of the divergence-free field~\cite{Henley05a} and is known to be proportional to $\alpha$~\cite{Castelnovo11a}. Since the coarse-grained magnetization field of the FCSL is not divergence-free, one needs to be careful when defining the stiffness of the FCSL. However, since the divergence-full field is antiferromagnetic, it does contribute to the magnetization of the system. Magnetic fluctuations of the FCSL are thus expected to come from the divergence-free contribution of the magnetization and, according to our results, to be smaller than in spin ice. Indeed, with $\chi$ being the susceptibility, the variance of the magnetization $\Delta M=(\chi T)_{T\rightarrow 0}$ is 2.00(1) in spin ice~\cite{Isakov04a,Ryzhkin05a,Jaubert13a} and 1.52(2) in the FCSL~\cite{Brooks14a} with the same ratio as for the $\alpha$'s
\begin{eqnarray}
\frac{\Delta M_{\rm SI}}{\Delta M_{\rm FCSL}} = 1.30(2) = \frac{\alpha_{\rm FCSL}}{\alpha_{\rm SI}}
\label{eq:ratio}
\end{eqnarray}


\section{Energetic Coulomb potential from dipolar interactions}
\label{hole}

\indent From now on, we shall include long-range dipolar interactions as given in Eq.~(\ref{eq:dip}).

Starting from a FCSL configuration and flipping one spin creates a pair of topological defects. But as opposed to spin ice, these defects can be categorized into two different kinds. On one hand, if a minority spin is flipped, then a pair of 4 in / 4 out topological defects is created, breaking the local divergence-free contribution of the Helmholtz decomposition (see Fig.~\ref{fig:half}). On the other hand, if a majority spin is flipped, then a pair of 2 in - 2 out topological defects is created, breaking the local divergence-full contribution.

\begin{figure}
\centering\includegraphics[width=8cm]{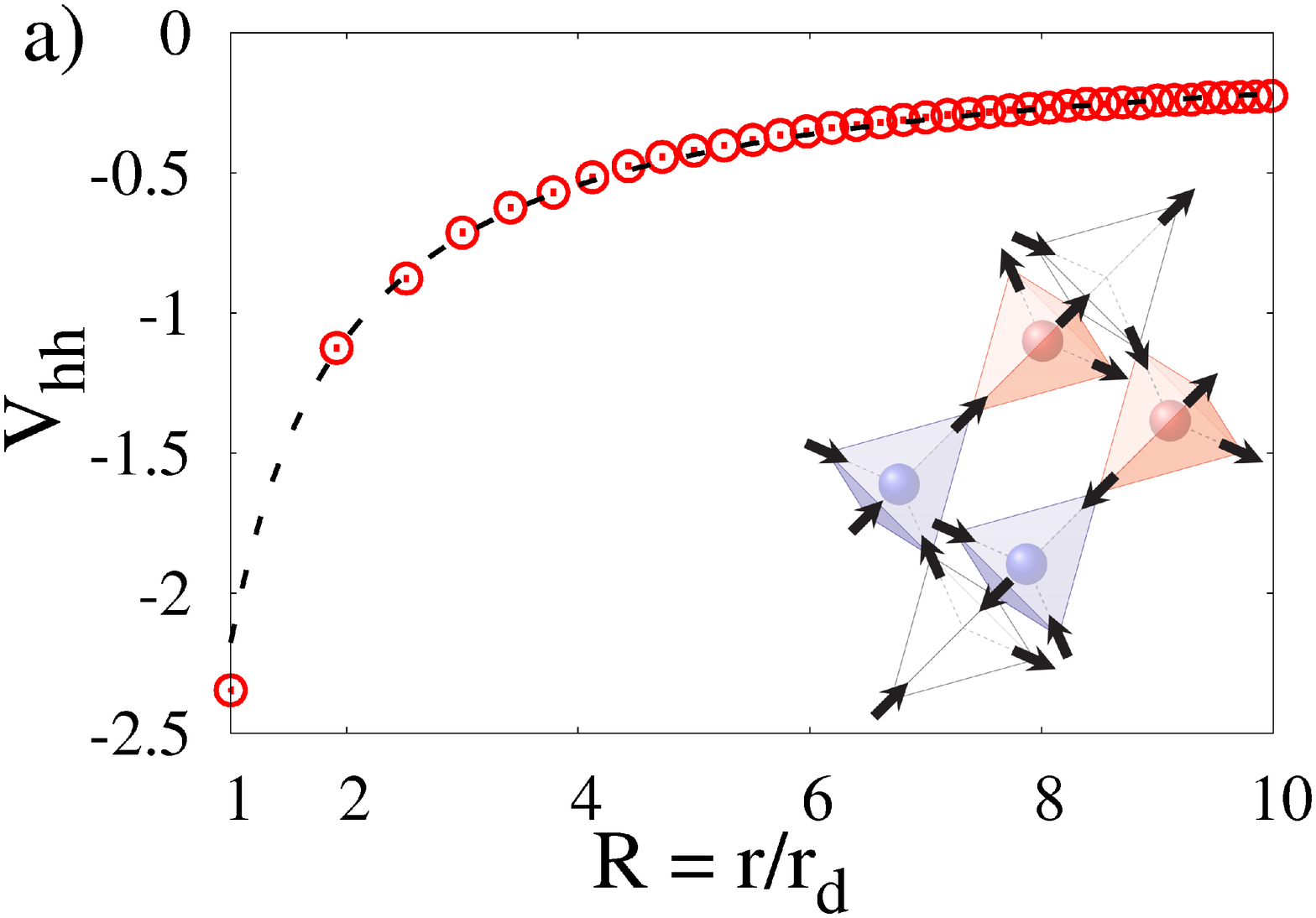}
\centering\includegraphics[width=8cm]{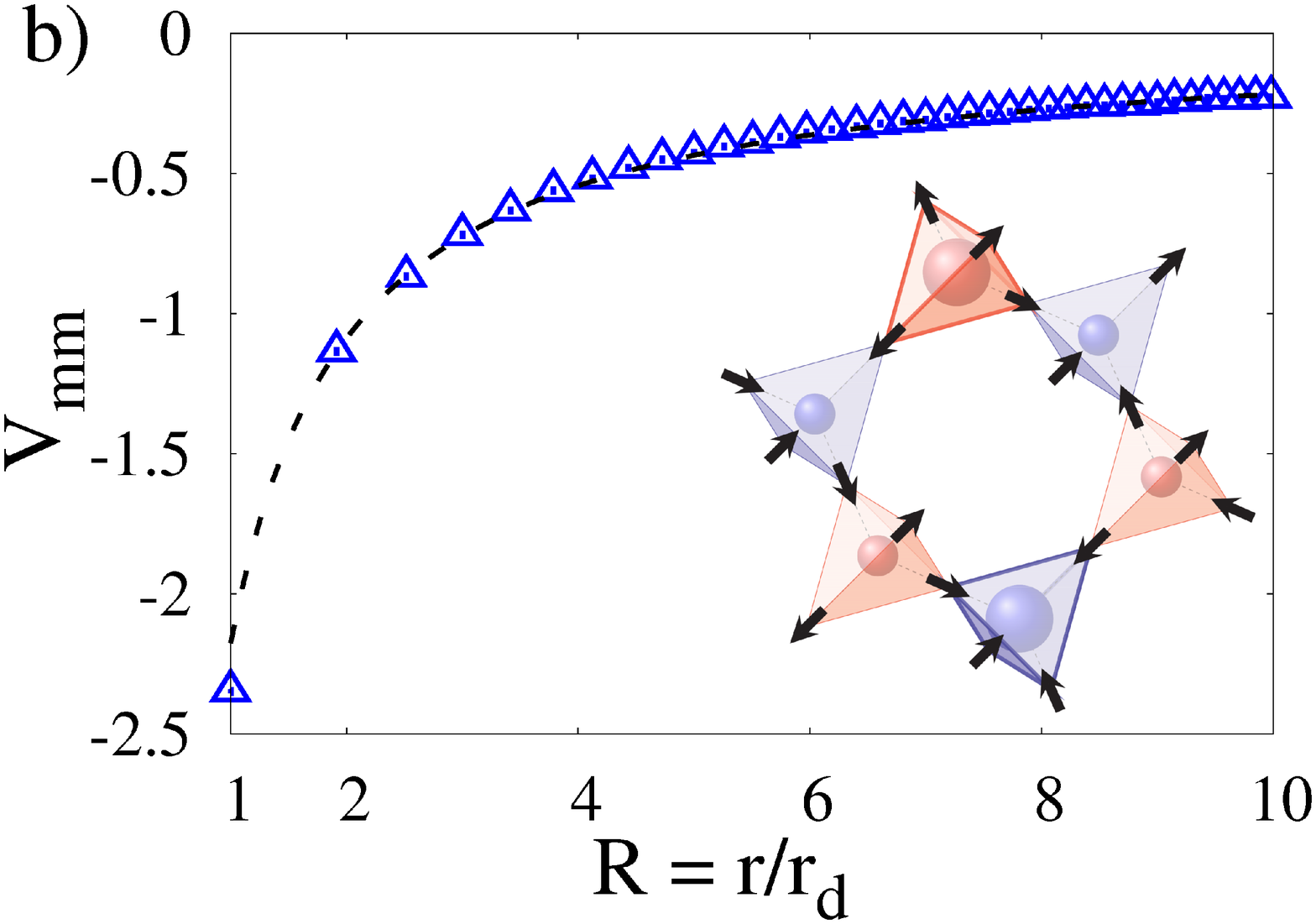}
\centering\includegraphics[width=8cm]{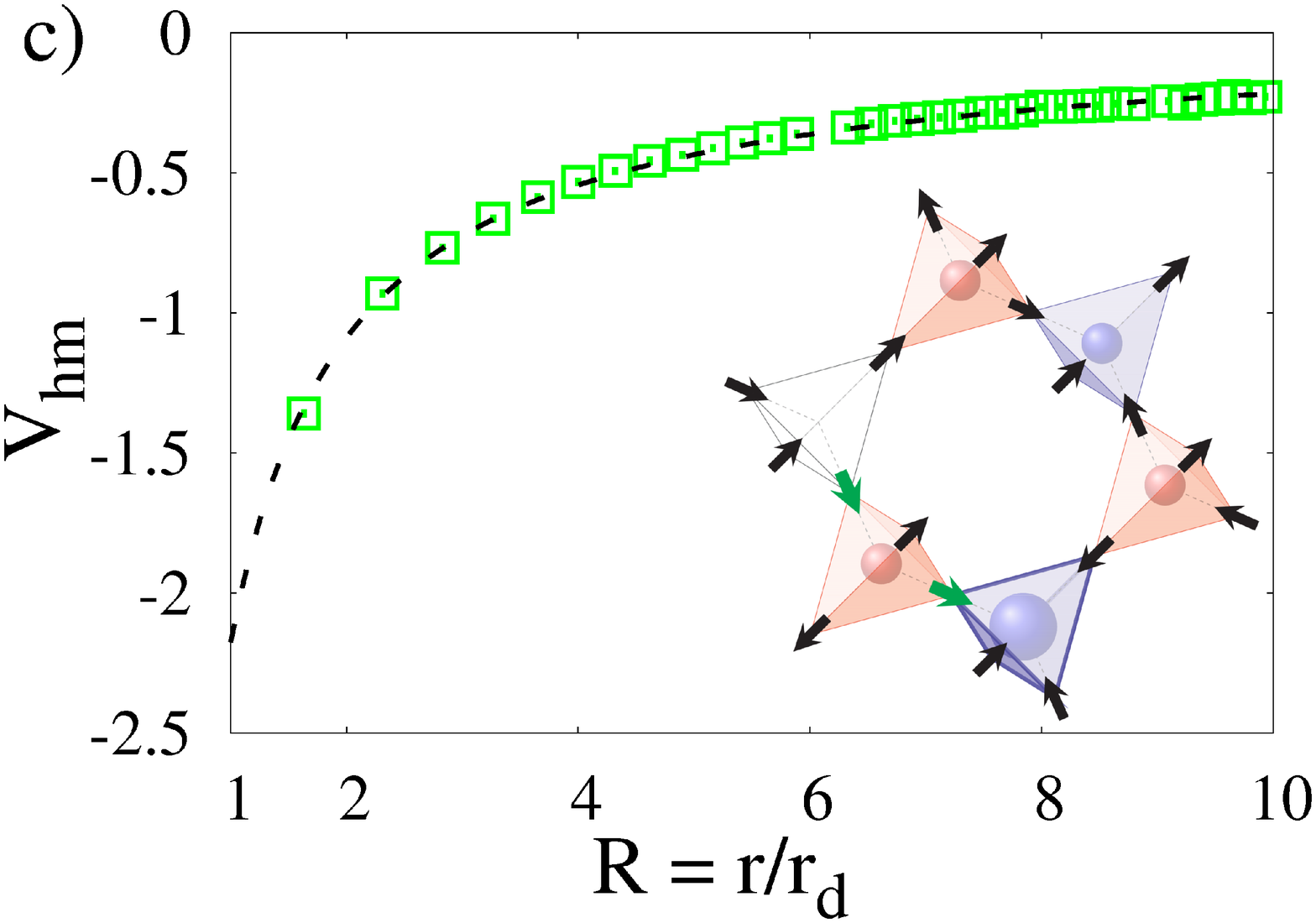}
\caption{Averaged potential energy between a) two ``2 in - 2 out'' defects (\textcolor{red}{$\circ$}), b) ``4 in'' and ``4 out'' defects (\textcolor{blue}{$\triangle$}), c) ``2 in - 2 out'' and ``4 in / 4 out'' defects (\textcolor{green}{$\square$}), as illustrated in the inset of each panel, obtained from Monte Carlo simulations with dipolar interactions. ``2 in - 2 out'' tetrahedra are colorless, while the ``4 in / 4 out'' tetrahedra are represented by thick lines. The green spins in panel c) represent the double-spin motion responsible for the creation of the ``2 in - 2 out'' and ``4 in / 4 out'' defects. To compare the simulations with Eq.~(\ref{eq:Vpot}) (dashed lines), the data have been shifted by a reference energy, serving as fitting parameter, whose origin is discussed in Eqs.~(\ref{eq:DEh}$-$\ref{eq:DEhm}). There are no multiplicative prefactors, meaning that all defects carry a unit of magnetic charge. The dipolar energy scale $D$ is set to 1. The error bars, computed from the results of 100 independent samples, are smaller than the data symbols.}
\label{fig:Vpot}
\end{figure}

To study the nature of these topological defects, the averaged potential energy between them has been computed using the same worm algorithm as in section~\ref{entropic}. The total dipolar energy of the system is computed at each step of the defect diffusion and stored as a function of the distance between the two defects, until they eventually annihilate each other, forming a new FCSL configuration. The procedure is repeated for 10$^{4}$ worm updates, and averaged over 100 independent initial configurations, in order to obtain good statistics. Simulations were done for reasonably big system sizes (16 000 sites) in order to minimize finite size effects. Indeed, when the pair of defects is separated by half the system size, the influence of the mirror images due to periodic boundary conditions is not negligible.  For a systematic analysis, the potential energy between two ``2 in - 2 out'' defects (noted $V_{hh}$), between ``4 in'' and ``4 out'' defects (noted $V_{mm}$) and  between ``2 in - 2 out'' and ``4 in / 4 out'' defects (noted $V_{hm}$) were computed separately (see the three panels of Fig.~\ref{fig:Vpot}).

Our results clearly confirm the emergent Coulomb potential between topological defects. It should be emphasized that this potential is directly coming from the dipolar energy between spins and should not be confused with the entropic potential considered in section~\ref{entropic}. Furthermore, as shown later in section~\ref{R}, the FCSL degeneracy is weakly lifted by dipolar interactions. Even if small when compared to the total energy of the system, this degeneracy lift is extensive. It is thus remarkable that despite extensive fluctuations of the total dipolar energy of the system, especially for configurations with finite magnetization, the averaged interaction between a single pair of defects is quantitatively described by the Coulomb potential of Eq.~(\ref{eq:Vpot}).\\

Based on the dumbbell model~\cite{Castelnovo08a}, it is possible to quantify these excitations analytically. When a pair of topological defects is created, the energy cost $\Delta E$ has three origins:
\begin{itemize}
\item the chemical potential of each defect, $p_{h}$ for ``2 in - 2 out'' and  $p_{m}$ for ``4 in / 4 out'';
\item the Coulomb attraction between the newly created pair of neighboring opposite charges, $V_{nn}=-\frac{8}{3}\sqrt{\frac{2}{3}}D$;
\item the Coulomb potential between each defect and the rest of the charge-ordered crystal, $V_{M}=-\frac{8}{3}\sqrt{\frac{2}{3}}D\,M_{\rm zb}$, where $M_{\rm zb}=1.638$ is the Madelung constant of the zinc-blende structure~\cite{Brooks14a}.
\end{itemize}
As computed in the Supplementary Information of Ref.~[\onlinecite{Castelnovo08a}] for spin ice, the dipolar interactions also contribute to the chemical potential of magnetic monopoles. When applied to the FCSL, the chemical potentials can be rewritten as
\begin{eqnarray}
p_{h}&=&\overline{p}_{h}\,+\,\frac{8}{3}\left(1+\sqrt{\frac{2}{3}}\right)D\label{eq:ph}\\
p_{m}&=&\overline{p}_{m}\,-\,8\left(1+\sqrt{\frac{2}{3}}\right)D,
\label{eq:pm}
\end{eqnarray}
where $\overline{p}_{i=\{h,m\}}$ is the chemical potential of non-dipolar origin, such as from Hamiltonian~(\ref{eq:ham}) for example. As discussed in section~\ref{FCSL}, the values of $\overline{p}_{i=\{h,m\}}$ are chosen such that the FCSL phase is stable. We now have all ingredients to estimate the energy cost for creating a pair of ``2 in - 2 out'' defects ($\Delta E_{hh}$) or ``4 in / 4 out'' defects ($\Delta E_{mm}$)
\begin{eqnarray}
\Delta E_{hh}&=&-2p_{h}\,+\,V_{nn}\,-\,2V_{M}\nonumber\\
&=&-2\overline{p}_{h}\,+\,\frac{16}{3}\left[-1+\left(M_{\rm zb}-\frac{3}{2}\right)\sqrt{\frac{2}{3}}\right]D\nonumber\\
&=&-2\overline{p}_{h}\,-\,4.73 D,\label{eq:DEh}\\
\Delta E_{mm}&=&-2p_{m}\,+\,V_{nn}\,+\,2V_{M}\nonumber\\
&=&-2\overline{p}_{m}\,+\,\left[16+\frac{8}{3}\sqrt{\frac{2}{3}}\left(5-2M_{\rm zb}\right)\right]D\nonumber\\
&=&-2\overline{p}_{m}\,+\,19.75 D.\label{eq:DEm}
\end{eqnarray}
In Monte Carlo simulations with dipolar interactions, corrections beyond the dumbbell model slightly modifies the numerical values of equations~(\ref{eq:DEh}) and~(\ref{eq:DEm}) and respectively give $-4.34 D$ and $+19.70 D$; the comparison is excellent for $\Delta E_{mm}$ and with a relative error of 8\% for $\Delta E_{hh}$.\\

As for spin ice~\cite{Castelnovo08a,Jaubert11c,Sala12a}, the emergent Coulomb potentials are thus rather natural in the context of the dumbbell model. But the motivation of this work is that the underlying magnetic order brings a new flavor to the problem. Once a pair of charges is created, additional spin flips separate the topological defects. Since the underlying long-range charge order takes place on a bipartite diamond lattice, with alternative positive and negative charges, a defect moving from one tetrahedron to the neighboring one alternatively becomes, for example, 2 in - 2 out, 4 in, 2 in - 2 out, 4 in ... carrying its topological and magnetic charge with it. This situation is analogous to kagome-ice physics~\cite{Moller09a,Chern11a}. Since ``2 in - 2 out'' and ``4 in'' defects have a priori different chemical potentials (see Eqs.~(\ref{eq:ph}) and~(\ref{eq:pm})), the hopping of topological defects from a tetrahedron to its neighbor alternatively costs and gains energy. Since the energy cost due to the chemical potential cancels out every two spin flips, this process does not confine the defects. Actually, this oscillating energy even disappears if we consider the simultaneous flipping of two spins \textit{successively pointing in and out}, ensuring that the hopping of the quasi-particle remains on the same diamond sublattice. We shall refer to such dynamics as a double-spin motion, which is reminiscent of a dimer move on the diamond lattice and of the mobility of holons and spinons in frustrated Mott insulators on bipartite lattices~\cite{Poilblanc11a}. The creation of a pair of ``2 in - 2 out'' and ``4 in / 4 out'' defects by a double-spin motion is depicted by green spins in the inset of Fig.~\ref{fig:Vpot}.c), and costs an energy
\begin{eqnarray}
\Delta E_{hm}&=&-p_{h}\,-\,p_{m}\,+\,V_{nnn}\nonumber\\
&=&-\overline{p}_{h}\,-\,\overline{p}_{m}\,+\,\left(4+\frac{16\sqrt{2}}{3\sqrt{3}}\right)D\nonumber\\
&=&-\overline{p}_{h}\,-\,\overline{p}_{m}\,+\,8.35D
\label{eq:DEhm}
\end{eqnarray}
where $V_{nnn}=-4D/3$ is the next-nearest-neighbour Coulomb potential energy (see Eq.~(\ref{eq:Vpot})). Please note that the Madelung term dissappears because the two defects with \textit{opposite} charges sit on the same diamond sublattice. A Monte Carlo average with Ewald summation gives $8.67D$, \textit{i.e.} a difference of 4\% with the numerical term of Eq.~(\ref{eq:DEhm}).

When created alone, a pair of ``4 in / 4 out'' defects only carries $\pm 1$ single charges (see Fig.~\ref{fig:Vpot}.b) as opposed to the $\pm 2$ double charges in spin ice. In the Helmholtz decomposition, this is because half of the zero-divergence has already been broken in the FCSL. However, when several topological defects are created, higher values of charges are possible. In the spin-ice model, the allowed magnetic charges are $\{-2,-1,0,+1,+2\}$. Out of the FCSL, they are $\{-3,-2,-1,0,+1,+2,+3\}$. If the long-range charge order of the FCSL imposes all up tetrahedra to be in a ``3 out - 1 in'' spin configuration, then a ``4 in'' defect would carry a $+3$ magnetic charge on up tetrahedra. This means that, as opposed to spin ice, the magnetic charge carried by a topological defect is now defined in the context of its environment, and in particular by the charge-order symmetry breaking of the FCSL. The consequence is that even if ``4 in / 4 out'' defects always have the same sign -- negative for 4 out, positive for 4 in -- the value of their charge is not uniquely defined.

As for a 2 in - 2 out defect, even the sign of its charge is unknown if considered alone. This is because a 2 in - 2 out defect breaks the local divergence-\textit{full} condition (see Fig.~\ref{fig:half}) and its sign is thus defined by the nature of the charge order that has been broken. If the charge order (\textit{i.e.} divergence-full condition) covers the entire system apart from some dilute topological defects, then the sign of the defect is rather trivially given by the up/down nature of the tetrahedron. However, if the system is covered by different domains of charge order -- \textit{i.e.} with different time-reversal symmetry breaking -- then the immediate vicinity of a given 2 in - 2 out defect is necessary to define the sign of its charge. The dynamics of the resulting domain walls, mediated by the co-existence with a Coulomb spin liquid and the emergence of topological magnetic charges would be a very interesting question to consider in the future.

\begin{figure}[t]
\centering\includegraphics[width=\columnwidth]{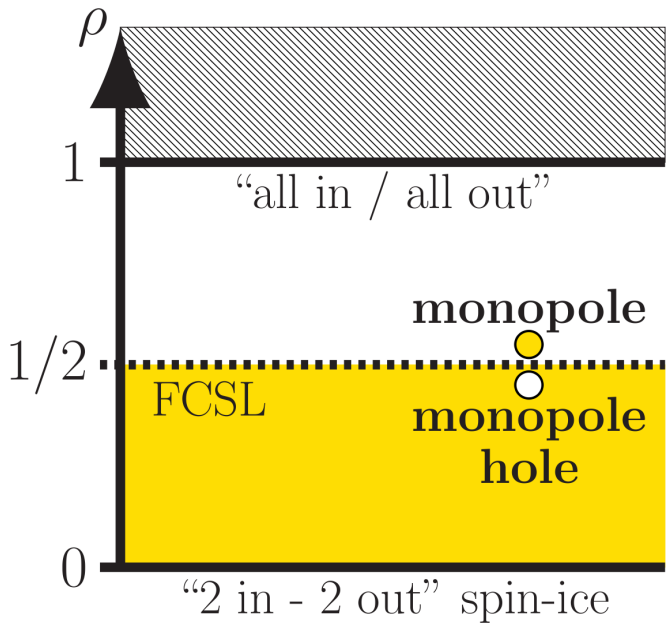}
\caption{
Schematic representation of the semiconductor analogy with the FCSL. The pseudo-magnetization $\rho$ of Eq.~(\ref{eq:sigma}) delimits the fraction of degrees of freedom respectively involved in the long-range ordered phase with divergence-full field (fully occupied valence band in yellow) and the Coulomb spin liquid with divergence-free field (empty conducting band). The divergence-free/full fields are explained in Fig.~\ref{fig:half}. A double-spin flip creates a \textit{gapped} excitation (see Eq.~(\ref{eq:DEhm}) and inset of Fig.~\ref{fig:Vpot}.c) taking the form of a pair of monopole and monopole hole which respectively break the local divergence-free and divergence-full fields. The gap is not visible in the figure because the monopole / monopole hole excitation conserves $\rho$. Further double-spin motions propagate these two quasi-particles within their respective conducting and valence bands, interacting only via an effective magnetic Coulomb potential (see Fig.~\ref{fig:Vpot}.c). In 3 dimensions, this potential is non-confining and gives rise to fractionalization as in semiconductors. As for the ``all in / all out'' phase,  the pseudo-magnetization is saturated at $\rho=1$. Topological defects thus necessarily take the form of monopole holes and are strongly confined since their diffusion requires the creation of further topological defects. In that sense, it is tempting to compare the pseudo-magnetization with the level of band filling: the hatched region represents an infinite gap because $\rho$ cannot be higher than 1, and the ``all in / all out'' phase is analogue to an insulator of magnetic charges.
}
\label{fig:fermi}
\end{figure}
\section{Semiconductor analogy}
\label{semiconductor}

\subsection{Monopole holes}

As a summary, a pair of defects created out of the FCSL by a double-spin motion, or dimer move, is made of a ``2 in - 2 out'' defect breaking the local divergence-full field, and a ``4 in'' or ``4 out'' defect breaking the local divergence-free field (see Fig.~\ref{fig:half} and inset of Fig.~\ref{fig:Vpot}.c). These defects carry a magnetic charge and interact via an emergent Coulomb potential (see Fig.~\ref{fig:Vpot}.c). The fractionalization of topological defects in spin ice has already been noticed to carry some analogy with semiconductors~\cite{Henley05a,Henley10a}. In the FCSL phase, we can go one step further. In semiconductors, an electron hole is defined by the lack of an electron from what should have been a fully-occupied valence band. From the Helmholtz decomposition, a ``2 in - 2 out'' defect can be understood as a \textit{monopole hole} in the otherwise charge-ordered phase, while the ``4 in / 4 out'' defect is a monopole in the same sense as in spin ice, breaking the local divergence-free condition. Please note that each defect may carry either a positive or negative magnetic charge. In absence of defects, the long-range order (full-divergence) and spin liquid (zero-divergence) respectively play the role of the fully occupied valence band and empty conducting band of semiconductors, with gapped excitations to create a pair of ``monopole / monopole hole'' (see Eq.~(\ref{eq:DEhm})). The $1/R$ potential ensures deconfinement of the topological defects and thus fractionalization of the excitations, even for monopole holes which break long-range order.

We should emphasize the importance of the double-spin motion in this analogy. Because of the time-reversal symmetry breaking of the long-range order, the divergence-full field of the FCSL is alternatively a source and a sink of fluxes on the bipartite diamond lattice. It means that if a negative magnetic charge is a monopole hole on up tetrahedra, it would be a monopole on down tetrahedra. Thus even if for a snapshot of the system at a given time $t$, the monopoles and monopole holes are properly defined, the nature of these defects is dynamically conserved only when hopping on the same diamond sublattice via double-spin motion. This is a direct consequence of the underlying dimer-model mapping. Of course here we are talking about Monte-Carlo dynamics. Real dynamics would be material dependent, but the system would likely take the form of a fluctuating ensemble of monopoles and monopole holes whose ratio would be determined by the temperature and the chemical potentials $p_{i=\{h,m\}}$, as if the doping was intrinsic rather than extrinsic.\\

An insightful way to appreciate the properties of the FCSL is to understand how it differs from the fully disordered (``2 in - 2 out'') and fully ordered (``all in / all out'') phases. To do so, the pseudo-magnetization $\rho$ is a useful observable, both globally for the entire system as defined in Eq.~(\ref{eq:sigma}) and locally at the level of each tetrahedron ($\rho_{\rm tet}=\left|\sum_{i\in \textrm{tet}} \sigma_{i}\right|/4$). In the ``2 in - 2 out'' spin-ice ground state, the absence of magnetic order with $\rho_{\rm tet}=0$ everywhere simply prevents the creation of monopole holes. This is why spin ice has often been discussed as a magnetic equivalent of an electrolyte~\cite{Bramwell09a,Castelnovo11a}.  Please note that both for spin ice~\cite{Jaubert09a,Castelnovo10d,Bramwell09a} and the FCSL, only \textit{alternating} currents of magnetic charges can subsist on long time scales. As for the ``all in / all out'' long-range order ($\rho_{\rm tet}=1$ everywhere), this phase has recently raised much interest in the context of pyrochlore Iridates Nd$_{2}$Ir$_{2}$O$_{7}$~\cite{Tomiyasu12a} and Tb$_{2}$Ir$_{2}$O$_{7}$~\cite{Lefrancois15a}. In comparison with the FCSL, ``all in / all out'' order means that the conducting band has been fully filled by divergence-full monopoles, making the pseudo-magnetization an effective measure of the level of band filling, as illustrated in Fig.~\ref{fig:fermi}. Interestingly, defects out of the ``all in / all out'' phase are necessarily \textit{confined} monopole holes with no empty band available since $\rho$ cannot be higher than 1. In that sense, the ``all in / all out'' phase is similar to an insulator of magnetic charges with an infinite gap. A possible follow-up of this analogy would be to study how this ``insulator'' could couple with a nearby Coulomb phase, such as for example in FeF$_{3}$~\cite{Sadeghi15a} or possibly with the presence of two interpenetrating magnetic pyrochlore lattices such as in pyrochlore Iridates.\\

The framework we have developed here provides a working example of the kind of emergent phenomena arising from the co-existence of magnetic order and spin liquids. The semiconductor analogy is able to catch a surprising number of characteristic features of the FCSL. There are of course limitations to this analogy, which we believe are good places to look for exotic physics, such as the possibility of domain walls separating domains with opposite time-reversal symmetry or the inherent topological nature of the FCSL which would probably lead to phase transitions beyond the standard Landau-Ginzburg-Wilson paradigm~\cite{Alet06a,Jaubert08a,Powell08b,Powell09a,Jaubert10a}.

\begin{figure}[h]
\centering\includegraphics[width=0.8\columnwidth]{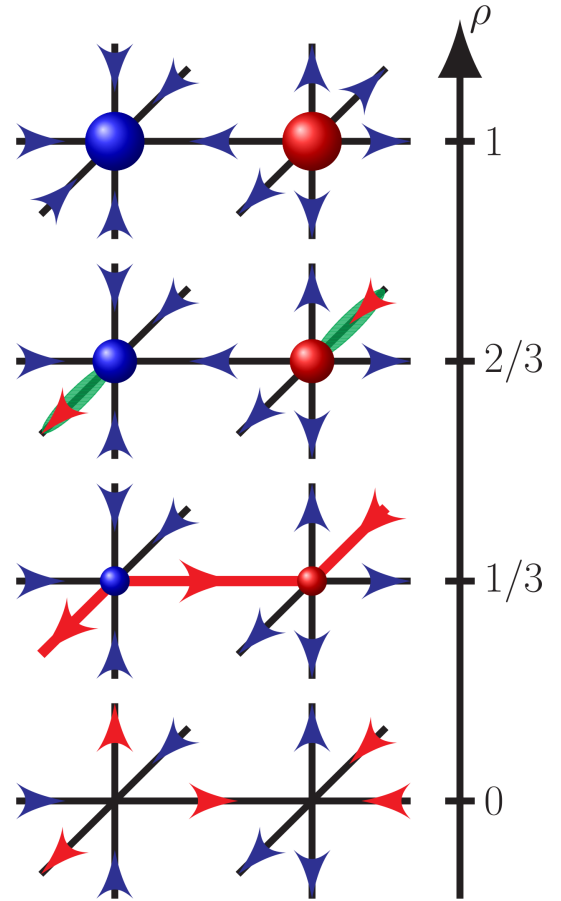}
\caption{
Vertices on the cubic lattice have 6-fold connectivity as opposed to the tetragonal geometry of the diamond lattice. This higher connectivity gives rise to two kinds of FCSL, \textit{i.e.} with co-existence of long-range order ($\rho>0$) and Coulomb phase: the hard-core dimer model at $\rho=2/3$ (dimers are in green) and the fully packed loop model at $\rho=1/3$ (the loop is a thick red line). Upon monopoles doping (\textit{i.e.} increasing $\rho$), it is thus possible to visit two phases similar to semiconductors before encountering the ``insulating'' all in / all out long-range order at $\rho=1$.
}
\label{fig:cubic}
\end{figure}

\subsection{Beyond pyrochlores}
\label{cubic}

So far, we have focused on the pyrochlore lattice in order to develop a comprehensive background for our theory. But the present analogy between semiconductor physics and a partially ordered Coulomb phase is not restricted to the pyrochlore symmetry. The microscopic mechanism able to stabilize a FCSL would always remain a model-dependent issue, but conceptually, our theory can be easily extended to most vertex models supporting a Coulomb phase. A straightforward ramification would be the bipartite hexagonal ($Ih$) ice lattice, which is the standard form of water ice at ambient pressure~\cite{Pauling35a} and of recent interest for quantum proton dynamics~\cite{Ryzhkin99a,Bove09a,Benton15b,Isakov15a}. At lower dimensions, the same analogy can be made on the checkerboard lattice but with the caveats that entropic correlations are logarithmic in two dimensions, while magnetic dipolar interactions are known to lift the ``2 in - 2 out'' degeneracy at the vertex level, a common problem in artificial spin ice~\cite{Wang06a,Moller06a}. The latter issue is absent in kagome ice, and so is the equivalent ``2 in - 2 out'' phase since a proper definition of the divergence-free field on kagome requires time-reversal symmetry breaking. Interestingly, a FCSL is known to be stable over a finite temperature window in presence of dipolar interactions on kagome~\cite{Moller09a,Chern11a}, or equivalently on the [111] magnetization plateau of spin ice materials~\cite{Matsuhira02a,Moessner03a}.\\

As illustrated by kagome, the connectivity of the vertex does not need to be tetragonal. Let us thus consider the bipartite cubic lattice which has been recently studied in the context of unconventional phase transitions in dimer/monomer and loop models~\cite{Sreejith14a,Nahum11a} and emergent loop models in perovskite oxynitrides~\cite{Camp12a}.We do not consider long-range dipolar interactions here, and thus the defects do not carry any effective magnetic charges. We impose that each bond of the cubic lattice carries a magnetic flux represented by an arrow towards one of the two neighbouring vertices. As opposed to the diamond lattice, each vertex has 6 legs (or arrows). By imposing all vertices of a given sublattice to have the same number of entering fluxes (inwards arrows), one can define various Coulomb phases as depicted in Fig.~\ref{fig:cubic}:
\begin{itemize}
\item 3 entering fluxes: there is no long-range order ($\rho=0$).
\item 2 entering fluxes: there is partial order ($\rho=1/3$) and the phase is mapped onto the fully packed loop model without loop crossing. Fluctuations of the loops represent fluctuations of the co-existing Coulomb spin liquid.
\item 1 entering flux: there is partial order ($\rho=2/3$) and the phase is mapped onto the hard-core dimer model.
\item 0 entering flux: the system is fully ordered ($\rho=1$).
\end{itemize}

On the pyrochlore lattice, when starting from the FCSL ($\rho=1/2$) and filling the conducting band with ``standard'' monopoles, one would immediately reach the insulating ``all in / all out'' order ($\rho=1$). On the cubic lattice on the other hand, starting from $\rho=1/3$ and filling the conducting band, one would first reach another FCSL with $\rho=2/3$, offering the tempting analogy with multi-band physics. 

\section{Ground state of the dipolar FCSL: maximization of flippable plaquettes}
\label{R}

Even if the potential energy between a pair of defects obeys the Coulomb law on average (see Fig.~\ref{fig:Vpot}), the ensemble of configurations of the FCSL remains \textit{quasi}-degenerate. It means that at low enough temperature, the system should ultimately order.\\

\begin{figure}[h]
\centering\includegraphics[width=8cm]{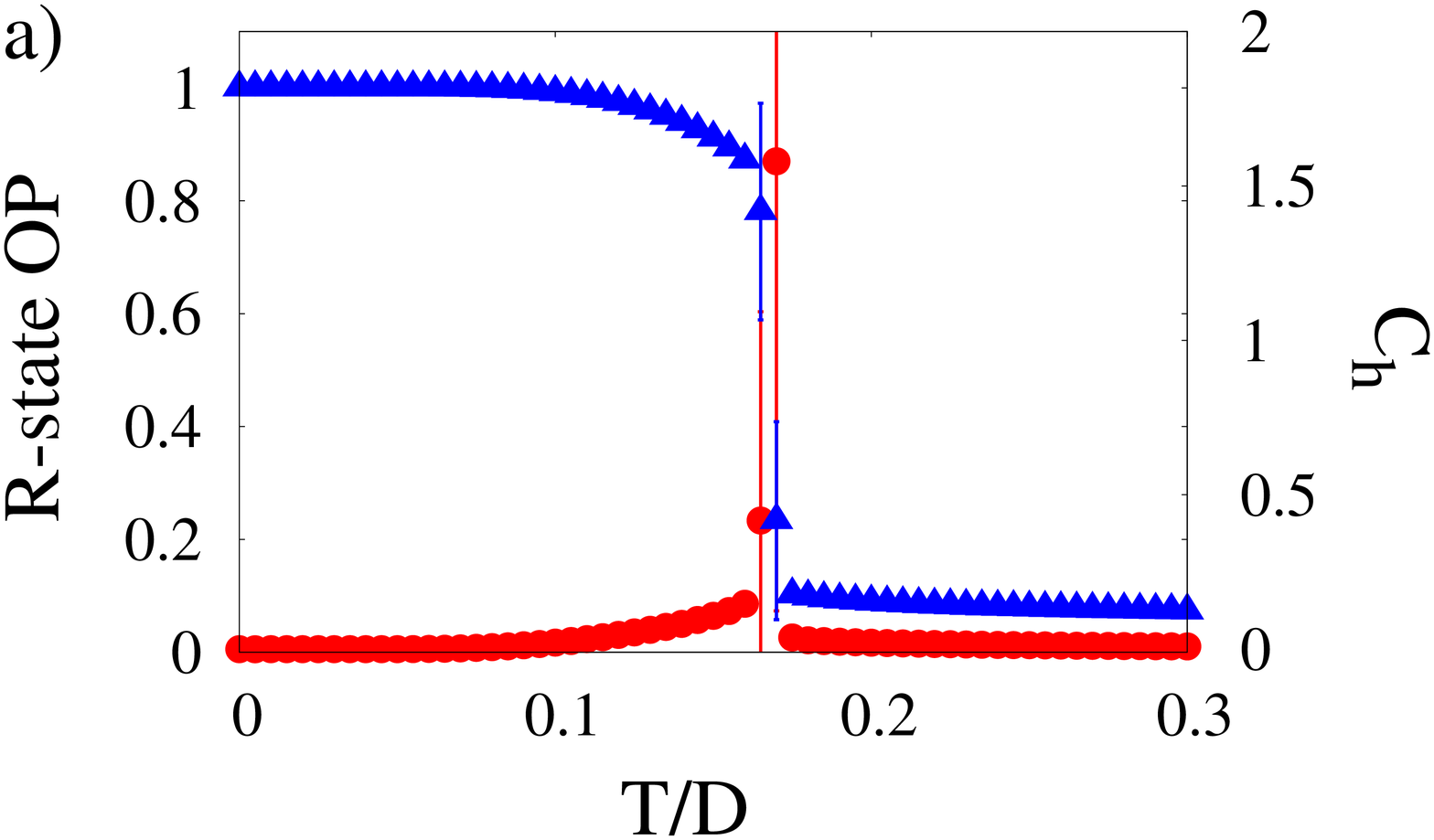}
\centering\includegraphics[width=8cm]{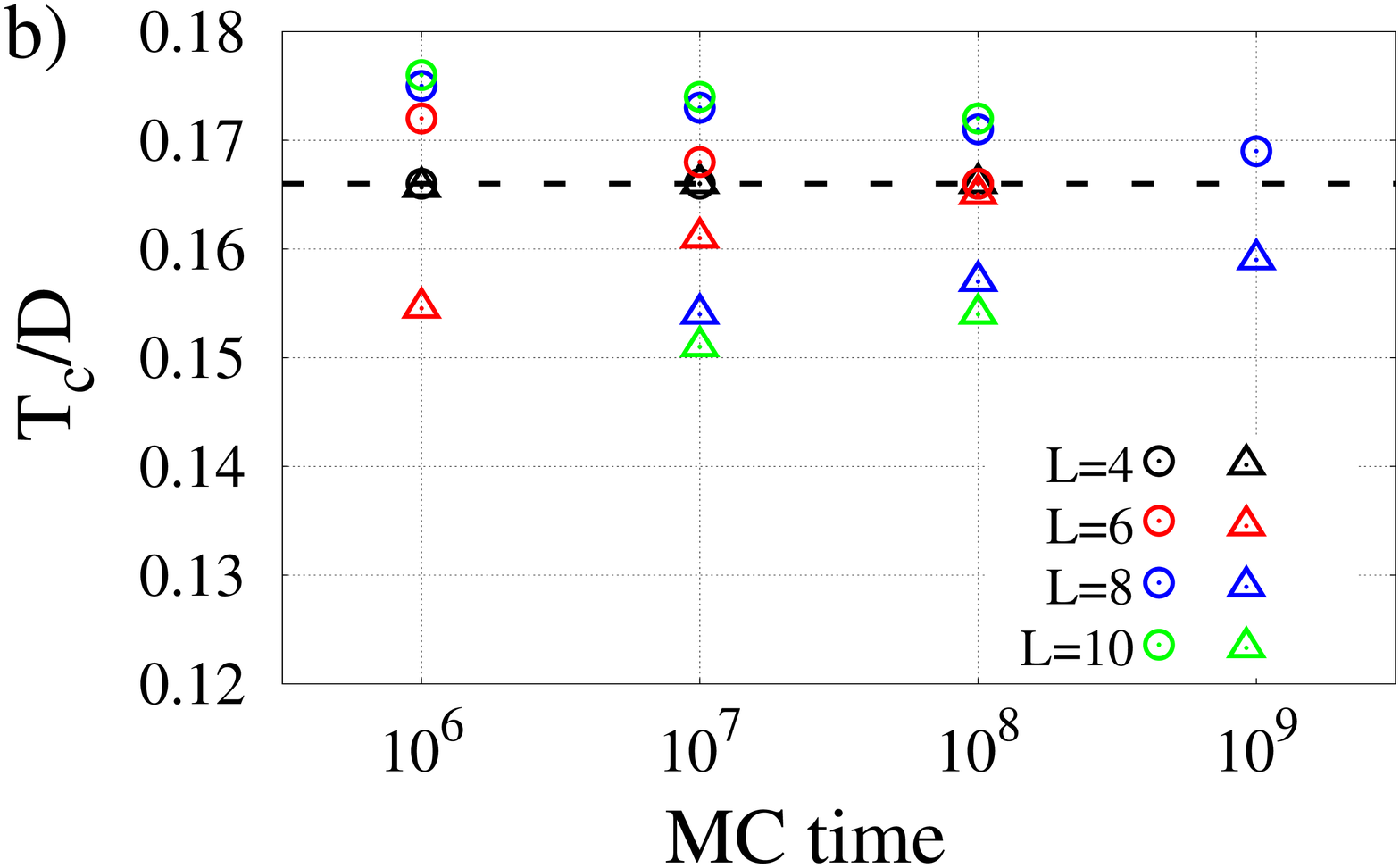}
\caption{At very low energy scale, dipolar interactions ultimately lift the FCSL degeneracy at $T_{c}/D=0.166\pm 0.003$. a) The first-order nature of the transition is clearly visible in the specific heat $C_{h}$ (\textcolor{red}{$\bullet$}) and in the discontinuity of the R-state order parameter (\textcolor{blue}{$\blacktriangle$}). The data are obtained from Monte Carlo (MC) simulations within the FCSL ensemble, averaged over 10$^{7}$ worm updates and 10 independent initial configurations, for a system size of 3456 pyrochlore sites. b) The hysteresis of the transition makes it difficult to precisely estimate the transition temperature. This is why in order to show convergence of the simulations, two kinds of equilibration processes were used. The system was either \textit{quenched} into one of the the R-states ($\circ$) or slowly annealed from high temperature ($\triangle$), which respectively over- and under-estimate the transition temperature. The MC time $t_{MC}$ of the $x-$axis represents the number of worm updates which are accepted/rejected based on a Metropolis argument accounting for the dipolar interactions. For both equilibration processes, the simulations were equilibrated at the temperature of measurements during $t_{MC}/10$. The results are given for different system sizes, showing convergence of the simulations towards $T_{c}/D=0.166\pm 0.003$.
}
\label{fig:Ch}
\end{figure}

\begin{figure}[h]
\centering\includegraphics[width=8cm]{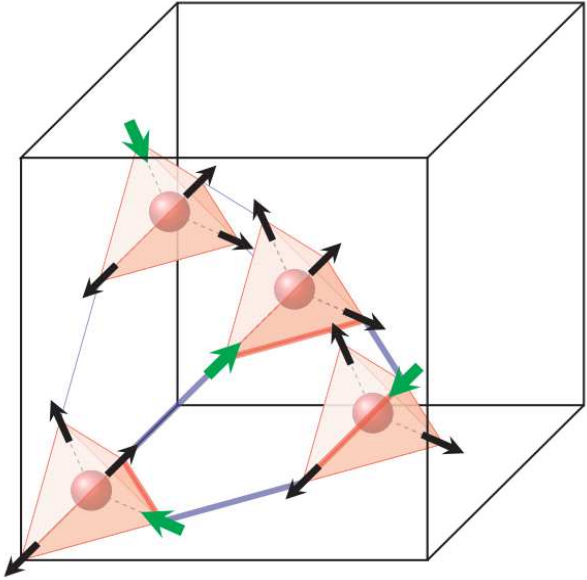}
\caption{
Unit cell of the eight-fold degenerate ground state of the FCSL with dipolar interactions, as observed in Ref.~[\onlinecite{Borzi13a}], which we identify as the R-states. Within a unit cell, the magnetization of the four down tetrahedra point in all four different local (111) directions. The R-state is known for maximizing the number of flippable plaquettes within the FCSL. The minority spins are colored in green and the flippable plaquette is shown by thick lines.
}
\label{fig:R}
\end{figure}

\begin{figure*}
\centering\includegraphics[width=16cm]{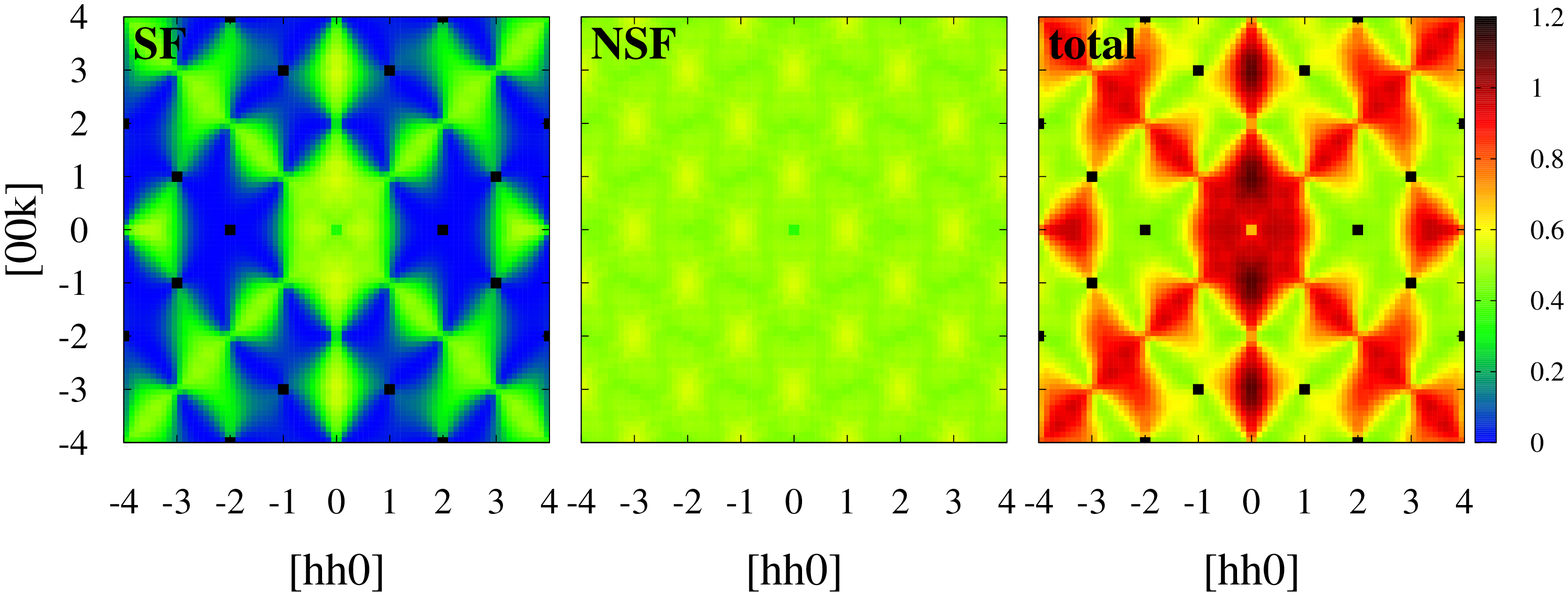}
\centering\includegraphics[width=16cm]{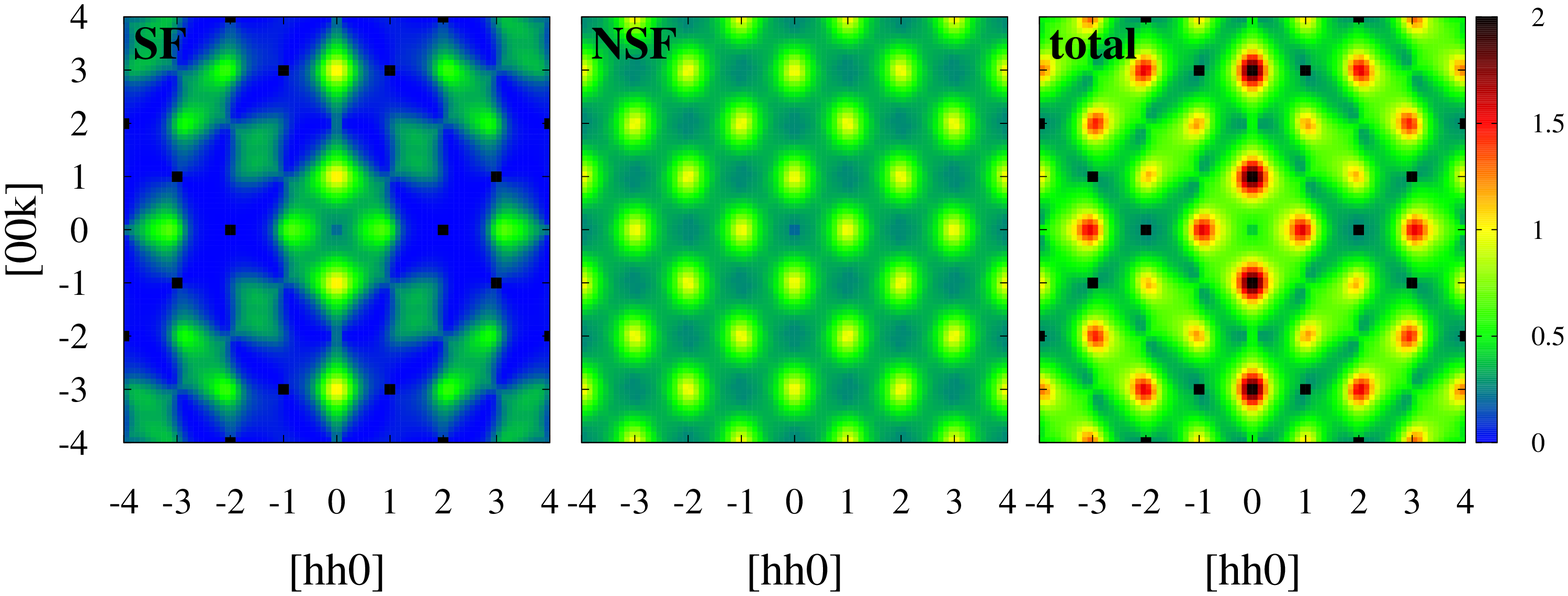}
\caption{
Structure factor of the FCSL with dipolar interactions obtained from Monte Carlo simulations at $T/D=2$ (\textit{top}) and $T/D=0.225$ (\textit{bottom}), showing the Spin-Flip (\textit{left}), Non-Spin-Flip (\textit{middle}) and total (\textit{right}) scattering. The NSF and SF channels measure the spin components along the $\vec e_{NSF}=(1\bar{1}0)$ and $\vec e_{SF}$ vectors respectively. $\vec e_{SF}$ is defined by forming an orthogonal basis with $\vec e_{NSF}$ and the wave-vector $\vec q$. Both temperatures are above the transition temperature of $T_{c}/D=0.166$. Black dots in the left panels are Bragg peaks signaling the long-range charge order of the FCSL~\cite{Brooks14a}. Despite the build-up of R-states correlations in the NSF channel, pinch points are still visible in the total scattering.
}
\label{fig:SQ}
\end{figure*}

As shown in Fig.~\ref{fig:Ch}, Monte Carlo simulations within the FCSL ensemble (without any topological defect) confirm the nature of the ground state found in Ref.~[\onlinecite{Borzi13a}], which we identify as the so-called R-states depicted in Fig.~\ref{fig:R}. The phase transition is of first order and occurs at $T_{c}/D=0.166\pm 0.003$. Coming from high temperature, the build up of correlations due to the R-states is clearly visible in the Non-Spin-Flip channel of the structure factor (see Fig.~\ref{fig:SQ}), while the inherent divergence-full long-range order of the FCSL is signaled by well-defined Bragg peaks (black dots) in the Spin-Flip channel~\cite{Brooks14a}. Remarkably, the pinch points due to the divergence-free Coulomb phase persist even just above the transition temperature.

To find the R-state here is quite interesting. Indeed, the R-state has been first studied to explain the magnetisation plateau of HgCr$_{2}$O$_{4}$ and CdCr$_{2}$O$_{4}$~\cite{Penc04a,Bergman06b} and has been shown to be the eight-fold degenerate ground state of the quantum dimer model on the diamond lattice in the limit of large negative potential energy~\cite{Bergman06a,Sikora09a,Sikora11a}. In the quantum dimer language, a negative potential energy favor configurations with flippable plaquettes, \textit{i.e.} where rings of 3 dimers can dynamically resonate around a hexagon of 6 sites on the diamond lattice. As explained in Fig.~\ref{fig:CSL}, the FCSL is the classical counterpart of this quantum dimer model (modulo time-reversal symmetry), where the minority spins are the dimers. In our spin language, a flippable plaquette is represented by thick lines in Fig.~\ref{fig:R}; flipping the 6 spins around this hexagon is the smallest fluctuation of the Coulomb spin liquid keeping the long-range charge order intact, and is an excitation out of the R-states.\\

Our physical understanding of this analogy is that dipolar interactions, by definition, enforce the closing of magnetic flux lines. As a textbook example, dipolar interactions are the reason for Weiss domains in standard ferromagnets. Since these ``flippable plaquettes'' are the smallest ensemble of spins able to close a magnetic flux line, it is understandable that dipolar interactions may want to maximize the number of such plaquettes. At the classical level, this is similar to a negative potential energy in the quantum dimer model, which is why the R-states are found as the ground states of both models.

But this result raises a new question. The ground state of spin ice also has a quantum counterpart, which has been studied in terms of potential and kinetic energies of flippable plaquettes~\cite{Hermele04a,Banerjee08a,Benton12a,Kato15a}. But while dipolar spin ice orders in a $q=(001)$ antiferromagnetic ground state~\cite{Hertog00a}, its quantum counterpart orders into the so-called ``squiggle'' state~\cite{Shannon12a} for large negative potential energy. We propose that the reason of this difference comes from the long-range nature of the dipolar interactions, which favor the closing of the flux lines for all length scales. In particular, a finite macroscopic magnetization is strongly hindered by dipolar interactions. This is not the case for the potential term of the quantum model which is local. Hence the finite magnetization of the squiggle state along the (001) axis~\cite{Shannon12a} is forbidden by dipolar interactions, but not in the quantum model. Since the R-states are antiferromagnetic, we do not encounter this problem in the ordering of the FCSL.



\section{Conclusion}

The co-existence of partial magnetic order and spin fluctuations has been a recurrent experimental and often controversial observation over the past years, especially in rare-earth pyrochlores Yb$_{2}$Ti$_{2}$O$_{7}$~\cite{Hodges02a,Chang14a,Yaouanc13b,Lhotel14a}, Tb$_{2}$Ti$_{2}$O$_{7}$~\cite{Gardner03a,Fritsch14a}, Er$_{2}$Sn$_{2}$O$_{7}$~\cite{Lago05a,Sarte11a,Guitteny13b,Yan13a} and Vesignieite kagome~\cite{Quilliam11b,Colman11a,Yoshida13a}. Here we have developed a theoretical framework to characterize the co-existence of long-range order and a Coulomb gauge field which we have named a Fragmented Coulomb Spin Liquid.

As for excitations out of the spin-ice ground state, topological defects of the FCSL interact via effective entropic and energetic Coulomb interactions, the latter one being due to magnetic dipolar interactions between spins. Following the Helmholtz decomposition, topological defects can be categorized into two kinds: monopoles which break the local divergence-free field as in spin ice, and monopole holes which break the divergence-full field and are thus directly related to the co-existence with long-range order. Our work i) shows that partial long-range order does not necessarily prevent deconfinement of fractionalized excitations, which would be consistent with persistent dynamics, and ii) shed a new light on the monopole picture in analogy with semiconductors, as a direct consequence of the underlying long-range order. Being based on the simple decomposition of degrees of freedom between an ordered contribution and a fluctuating one, we expect this framework to be relevant to a wide range of systems (artificial spin ice, magnetisation plateaux...).

Finally, we have identified that dipolar interactions on the classical FCSL support the same ground state as large negative potential energy on the equivalent quantum dimer model~\cite{Bergman06a,Sikora09a,Sikora11a,Borzi13a}. This offers a natural classical mechanism to maximize the number of flippable plaquettes, which can be expected to be a general tendency of dipolar interactions, but not a systematic outcome; among other points, it depends on the long-distance geometric constraints.\\

Future directions of work would be to consider Heisenberg spins instead of Ising, which would for example allow for gapless excitations. At the classical level, the decomposition of degrees of freedom would be straightforward in terms of irreducible representations~\cite{Yan13a}: the divergence-full and divergence-free contributions would respectively correspond to \textsf{A}$_{2}$ and \textsf{T}$_{1}$ symmetries. Also, 
instead of one long-range antiferromagnetic phase, we could consider the influence of frozen disorder~\cite{Bellier-castella01a,Saunders07a,Shinaoka11a}, or nematic phases where quadrupolar decomposition might be necessary.

\begin{acknowledgments}
The author is thankful to Marion Brooks-Bartlett, Simon Banks, Laurent de Forges de Parny and Peter Holdsworth for collaborations on related topics and to Owen Benton, Andrew Smerald and Roderich Moessner for insightful comments on the manuscript. This work was supported by the Okinawa Institute of Science and Technology Graduate University.
\end{acknowledgments}
\bibliography{/Users/Ludo/Desktop/biblio}
\end{document}